\documentclass[twoside,slac_one]{revtex4}
\usepackage{graphicx}
\usepackage{fancyhdr}
\usepackage{amsmath} 
\usepackage{bm}
\usepackage{amsxtra}
\usepackage{amssymb}
\usepackage{amsthm}
\usepackage{latexsym}
\usepackage{lscape}

\begin{document}

\title{Search for the Standard Model Higgs boson in $WH \rightarrow l\nu bb$ and $H \rightarrow WW^{(*)} \rightarrow l\nu l\nu$ channels at ATLAS}

%

\author{Lashkar Kashif on behalf of ATLAS}
\affiliation{Department of Physics, University of Wisconsin, Madison, WI, USA}

\begin{abstract}

Results for the Standard Model Higgs boson search by the ATLAS experiment in the $WH \rightarrow l\nu bb$ and $H \rightarrow WW^{(*)} \rightarrow l\nu l\nu$ channels are presented. The results are based on 1.04 fb$^\mathrm{-1}$ of data from $p\bar{p}$ collisions at $\sqrt{s}$ = 7 TeV produced by the LHC in 2011. No evidence is found for the Standard Model Higgs boson in either decay mode. The $WH \rightarrow l\nu bb$ channel is not yet sensitive to the Standard Model Higgs, while the $H \rightarrow WW^{(*)} \rightarrow l\nu l\nu$ channel excludes the Standard Model Higgs in the range of 158 $<~m_\mathrm{H}~<$ 186 GeV at the 95\% confidence level.

\end{abstract}

\maketitle

\thispagestyle{fancy}

\section{Introduction}

$WH \rightarrow l\nu bb$ and $H \rightarrow WW^{(*)} \rightarrow l\nu l\nu$ are two of the most important channels to search for the Standard Model (SM) Higgs boson, together covering the theoretically allowed Higgs mass range. In this talk, we present Higgs search results in these channels by the ATLAS experiment at the CERN Large Hadron Collider (LHC). 

\begin{figure}[ht]
 \centering
 \includegraphics[width=80mm]{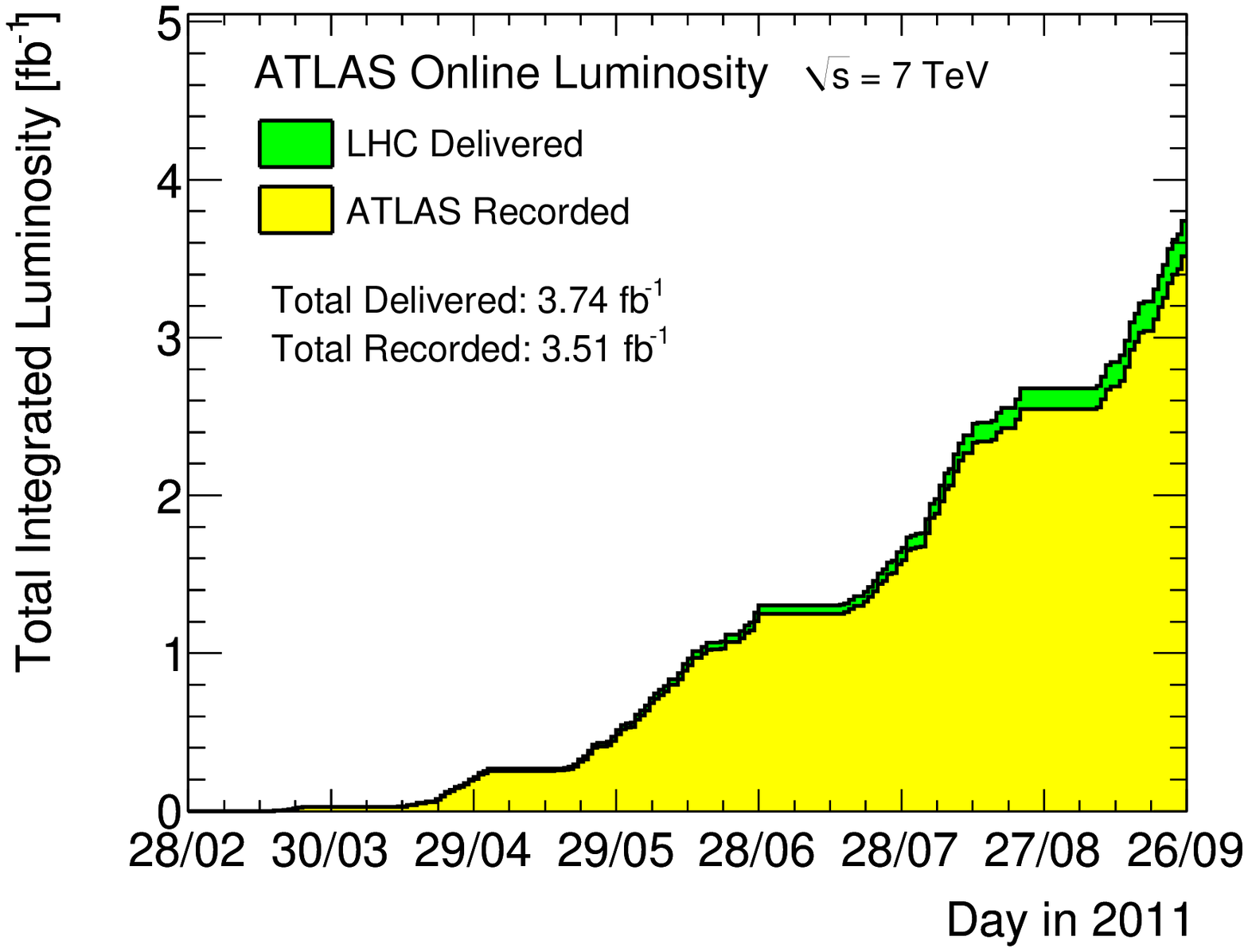}
 \includegraphics[width=80mm]{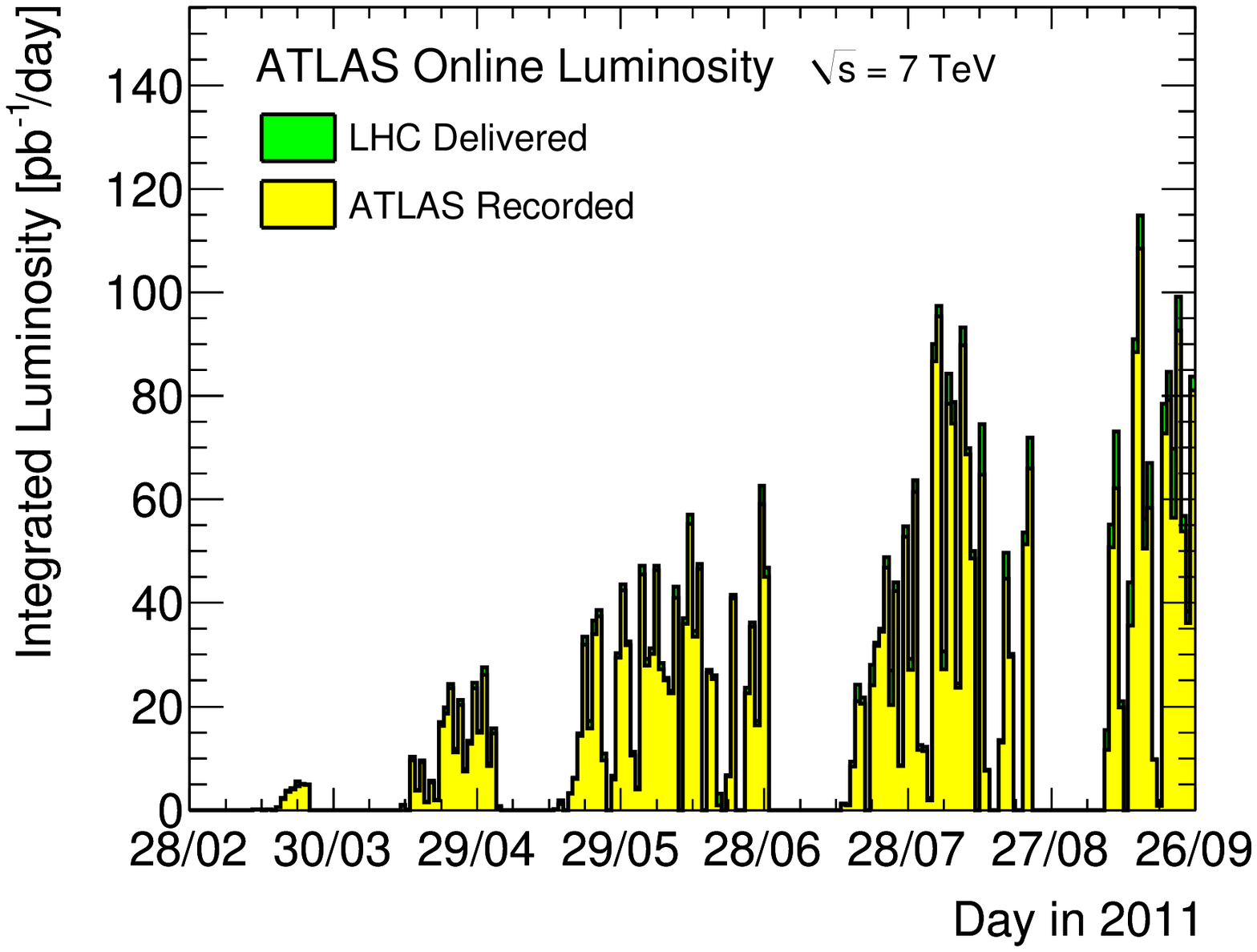}
 \caption{The integrated luminosity collected by the ATLAS experiment between March and August 2011 \textit{(left)}, and the instantaneous luminosity evolution during the same period\textit{(right)}.} 
 \label{fig:lumi_tot}
\end{figure}

The data sample used in these analyses was collected by ATLAS during the period March-June 2011, and corresponds to 1.04 fb$^\mathrm{-1}$ of integrated luminosity. Figure~\ref{fig:lumi_tot} shows the 7 TeV collision data accumulated by ATLAS in 2011 as function of time. At the time this article is written, the total recorded data correspond to 3.51 fb$^\mathrm{-1}$, of which less than one-third was used for the results reported herein. The uncertainty on the luminosity measurement is 3.7\%. 

Figure~\ref{fig:lumi_tot} also shows the evolution of the instantaneous luminosity $L_\mathrm{inst}$ as a function of time in 2011. For the results that follow, the maximum $L_\mathrm{inst}$ reached was 1.26 $\times~10^\mathrm{33}$ cm$^\mathrm{-2}$ s$^\mathrm{-1}$. The average number of proton-proton interactions per bunch crossing was about 6.

\section{The $WH \rightarrow l\nu bb$ channel}
\label{sec:wh}

\begin{figure}[ht]
 \centering
 \includegraphics[width=80mm]{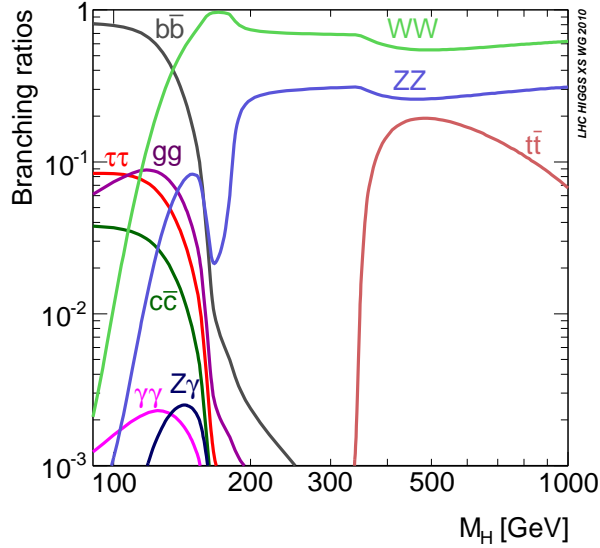}
 \caption{The branching ratio of the Standard Model Higgs boson into various final states as a function of the Higgs mass.} 
 \label{fig:higgs_BR}
\end{figure}

The $WH \rightarrow l\nu bb$ channel takes advantage of the large branching ratio of the $H \rightarrow bb$ decay, which is about 2/3 at the Higgs mass $M_\mathrm{H}$ of 120 GeV (Figure~\ref{fig:higgs_BR}). Owing to the very large background of prompt $b\bar{b}$ pairs, a direct search for the $H \rightarrow bb$ decay is extremely difficult at the LHC. However, Higgs production associated with a leptonically decaying $W$ or $Z$ boson makes the search easier, since the lepton can be triggered on, and the lepton requirement can be used to reduce the large $b\bar{b}$ background. As a reference, the cross-section $\times$ branching ratio of the process $WH \rightarrow l\nu bb$ is about 91 fb for $M_\mathrm{H}$ = 120 GeV.

A cut-based analysis is performed and its strategy is simple: we require two $b$-tagged jets in addition to $W$ boson search criteria. A Higgs signal would show up as an excess in a $b\bar{b}$ invariant mass spectrum.

\subsection{Event selection}
\label{sec:wh_evsel}

\begin{figure}[ht]
 \centering
 \includegraphics[width=80mm]{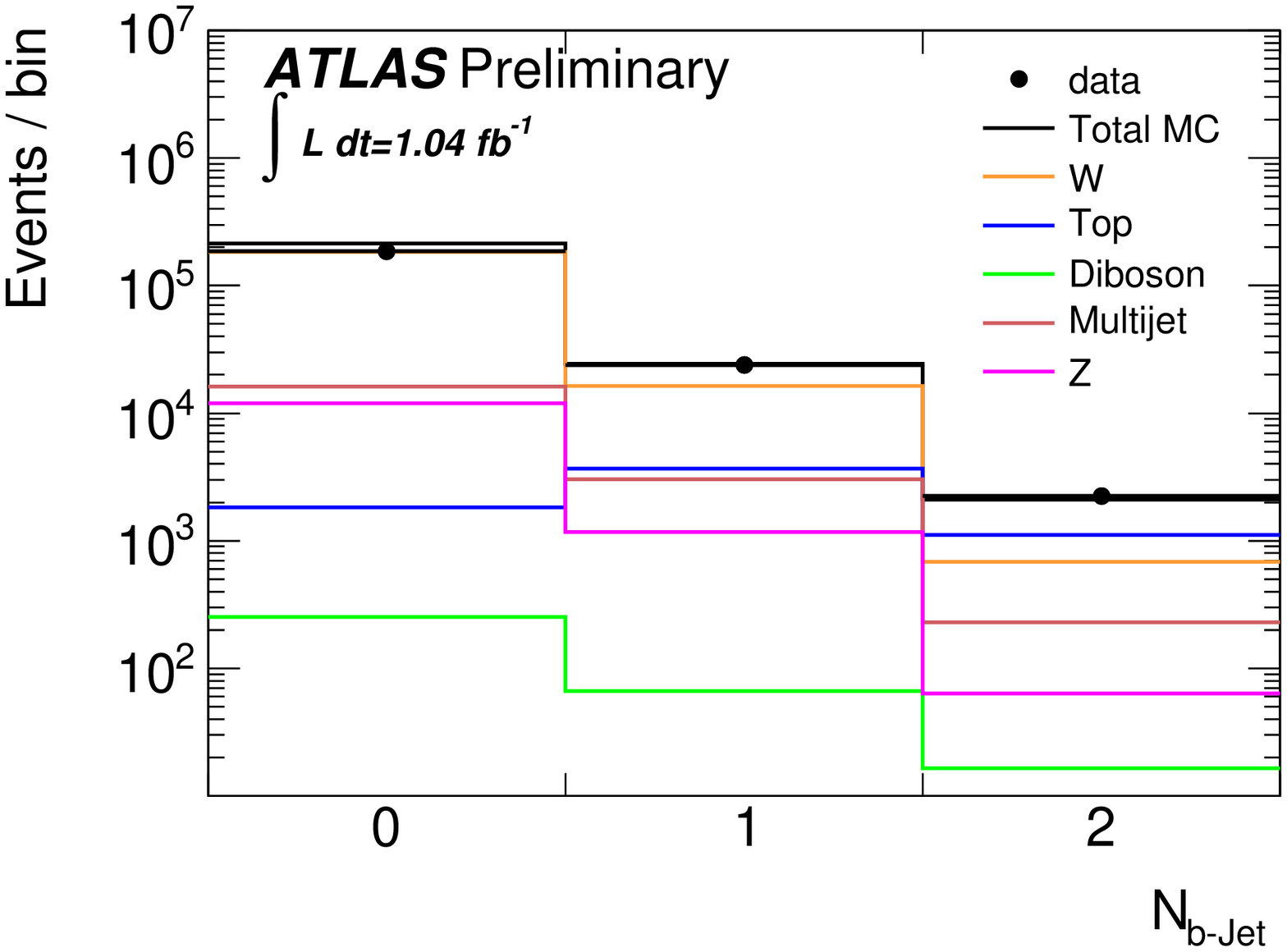}
 \includegraphics[width=80mm]{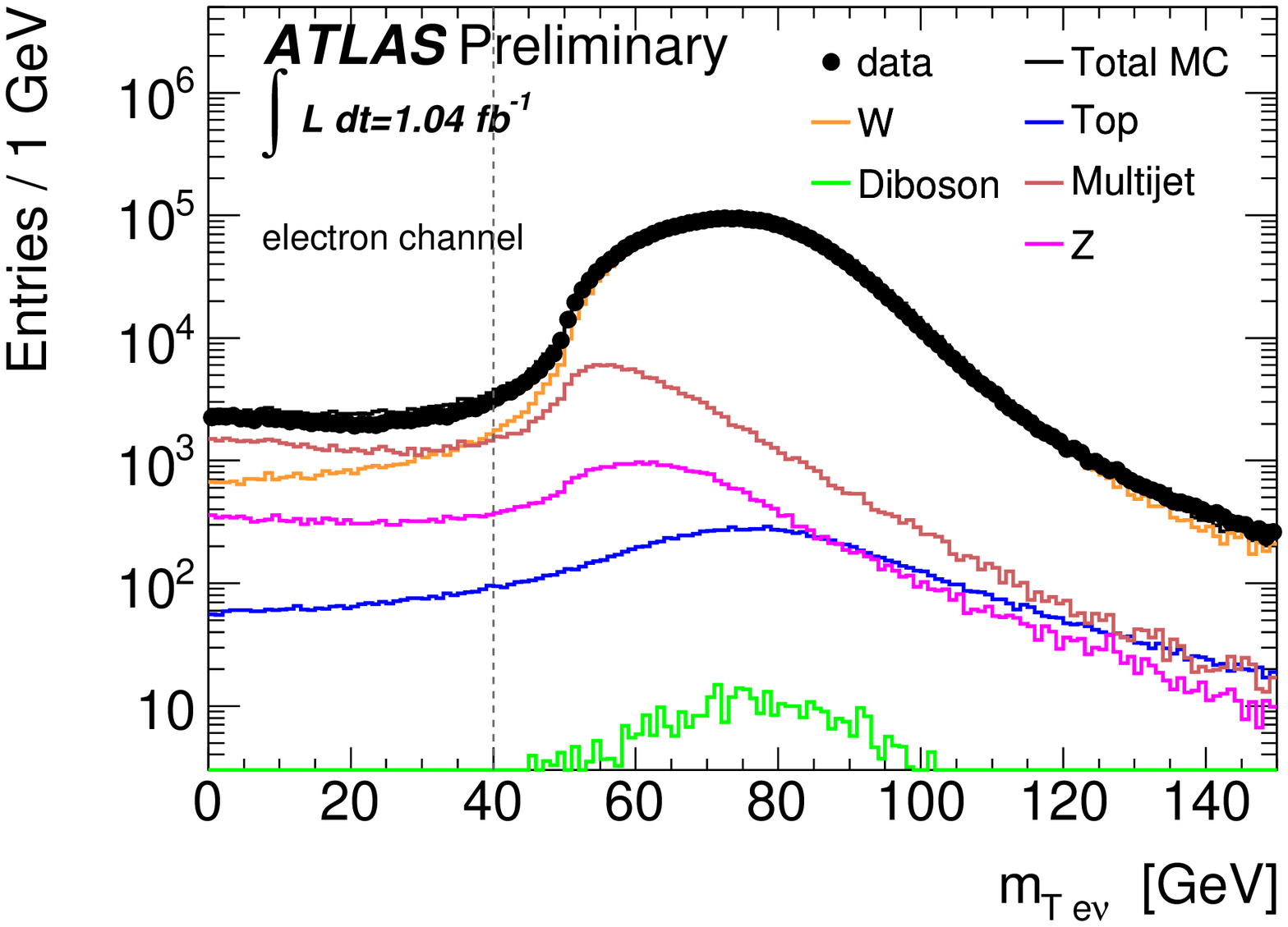}
 \caption{\textit{Left}: Distribution of the number of $b$-tagged jets per event after the requirement of exactly two jets in the event; \textit{right}: transverse mass spectrum in the electron channel after the cut on $E_\mathrm{T}^\mathrm{miss}$.}
 \label{fig:Nbjet}
\end{figure}

Since the $W$ boson can decay into an electron or a muon, the analysis is performed separately for the electron and muon channels. In each channel, we require that events pass the lowest-threshold unprescaled trigger: this threshold is 20 GeV and 18 GeV for the electron and muon channels respectively. Further event selection criteria are listed below:

\begin{itemize}

 \item Exactly one reconstructed lepton in the event. The lepton must have at least 25 GeV transverse momentum ($p_\mathrm{T}$). An electron must be within the pseudorapidity range $|\eta|~<~2.47$; a muon must be within $|\eta|~<~2.4$. 
 
 \item Exactly two jets in the event, reconstructed using the anti-k$_\mathrm{T}$ algorithm with a distance parameter R = 0.4. Each jet must have $p_\mathrm{T}~>~25$ GeV and be within $|\eta|~<~2.5$. In addition, each jet must be tagged as a $b$-jet. Figure~\ref{fig:Nbjet} \textit{(left)} shows the number of $b$-tagged jets per event.
 
 \item The missing transverse energy, $E_\mathrm{T}^\mathrm{miss}$, in the event must be larger than 25 GeV.
 
 \item The $W$ transverse mass $m_\mathrm{T}$ must be larger than 40 GeV, where $m_\mathrm{T}$ is defined as $m_\mathrm{T} = \sqrt{2p_\mathrm{T}^\mathrm{l}p_\mathrm{T}^\mathrm{\nu} (1-\cos(\phi^\mathrm{l}-\phi^\mathrm{\nu}))}$.  $p_\mathrm{T}^\mathrm{l}$ is the lepton $p_\mathrm{T}$,  $p_\mathrm{T}^\mathrm{\nu} = E_\mathrm{T}^\mathrm{miss}$, and $\phi$ is the azimuthal angle. Figure~\ref{fig:Nbjet} \textit{(right)} shows the transverse mass distribution in the electron channel.
  
 \end{itemize}
 
\subsection{Background processes}

\textit{Top}: $t\bar{t}$ and single top production constitute the largest background to the $WH \rightarrow l\nu bb$ channel. These backgrounds are largely reduced by the requirement of exactly two jets in the event. The shape of the residual top background is estimated from Monte Carlo (MC), but the normalization is extracted from a sideband fit in the di-$b$jet invariant mass $m_\mathrm{b\bar{b}}$ distribution in data. The sidebands are defined as $m_\mathrm{b\bar{b}}~<$ 80 GeV and 140 $<~m_\mathrm{b\bar{b}}~<$ 250 GeV. 

\begin{figure}[ht]
 \centering
 \includegraphics[width=80mm]{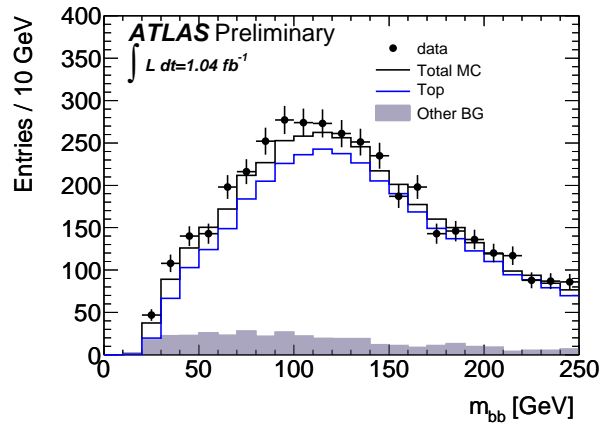}
 \caption{Di-$b$jet invariant mass $m_\mathrm{b\bar{b}}$ spectrum in the top control region in data, in top Monte Carlo and in other Monte Carlo-derived backgrounds.}
 \label{fig:topctrl}
\end{figure}

The shape and normalization of the top background are cross-checked by defining a top control region in data defined by having three jets in the event, two of which must be $b$-tagged. Figure~\ref{fig:topctrl} shows the $m_\mathrm{b\bar{b}}$ distribution in this region in data, in top MC and in other backgrounds. We see that the MC gives a reasonable description of both the shape and the normalization seen in data.

\textit{W+jets}: $W+b\bar{b}$ events constitute a large background at a low $m_\mathrm{b\bar{b}}$. The shape of the $W+b\bar{b}$ background is modeled using the shape of the untagged di-jet invariant mass $m_\mathrm{jj}$ distribution. This modeling is justified by checking that the shape of the $m_\mathrm{jj}$ spectrum in untagged \textit{W+jets} events is consistent with that of the $m_\mathrm{b\bar{b}}$ spectrum in MC. The template is normalized using a fit to $m_\mathrm{b\bar{b}}$ sidebands as for the top background.

\textit{QCD}: In both electron and muon channels, the QCD background is estimated from data using template fits. QCD-enriched templates in data are obtained by requiring non-isolated leptons; the normalization is derived from fits to the $E_\mathrm{T}^\mathrm{miss}$ distribution. The process is cross-checked in a QCD control region in data obtained by reversing the $E_\mathrm{T}^\mathrm{miss}$ and $m_\mathrm{T}$ cuts.

\textit{Diboson}: Diboson backgrounds are expected to be very small, and are estimated from MC.

\subsection{Systematic uncertainties}

Table~\ref{tab:wh_syst} summarizes the systematic uncertainties from experimental sources as well as those in the Higgs cross-sections for $m_\mathrm{H}$ = 115 GeV and $m_\mathrm{H}$ = 130 GeV. We see that by far the largest source of systematic uncertainty is $b$-tagging. 

\begin{table}[htbp]
\begin{center}
\begin{tabular}{|l|c|c|}
\hline
Source of Uncertainty   & \multicolumn{2}{c|}{Effect on $WH \rightarrow l\nu bb$ signal} \\
	                   & $m_\mathrm{H}=115~$ GeV & $m_\mathrm{H}=130~$ GeV  \\
\hline
Electron Energy Scale           &  1\%     &   1\%     \\
Electron Energy Resolution      &  1\%     &   1\%	\\
Electron  Efficiency            &  1\%     &   1\%	 \\
Muon Momentum Resolution        &  4\%     &   1\%	 \\
Muon  Efficiency                &  1\%     &   1\%	  \\
Jet Energy Scale                &  1\%     &   3\%	\\
Jet Energy Resolution           &  1\%     &   1\%	\\
Missing Transverse Energy       &  2\%     &   3\%	\\
$b$-tagging Efficiency          &  16\%    &   17\%	\\
$b$-tagging Mis-tag Fraction    &  3\%     &   3\%	  \\
\hline                            
Luminosity                      &  4\%     &   4\%     \\
Higgs Cross-section             &  5\%     &   5\%	\\
\hline 
\end{tabular}
\end{center}
\caption{\label{tab:wh_syst} Effect of systematic uncertainties on signal yield for two different Higgs masses.}
\end{table}

\subsection{Results}

\begin{figure}[ht]
 \centering
 \includegraphics[width=80mm]{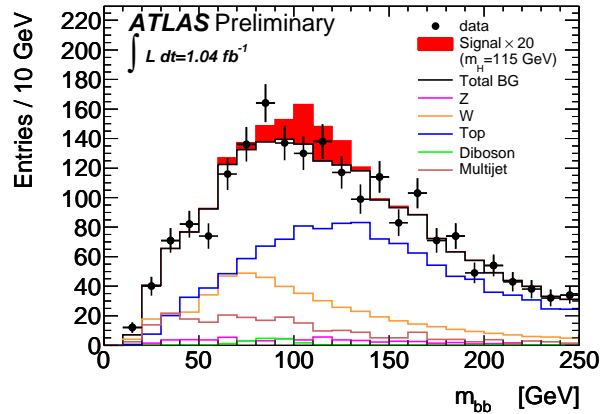}
 \caption{$m_\mathrm{b\bar{b}}$ distribution in data, for a 115 GeV Higgs signal and for the various backgrounds. The MC prediction for signal has been amplified by a factor of 20 for visibility.}
 \label{fig:wh_resbbmass}
\end{figure}

Table~\ref{tab:wh_res} lists the background events expected in 1.04 fb$^\mathrm{-1}$ of data, the number of data events actually observed and expected signal event count for five different Higgs masses in the range 110 $<~m_\mathrm{H}~<$ 130 GeV. The numbers were obtained in the range of 40 $<~m_\mathrm{b\bar{b}}~<$ 250 GeV. We see that the number of observed events is consistent with the predicted background. Figure~\ref{fig:wh_resbbmass} shows the $m_\mathrm{b\bar{b}}$ distribution observed in data as well as the expected distribution from MC, the signal prediction having been amplified by a factor of 20 for visibility. We do not see any excess in the observed distribution. We therefore conclude that we do not find any evidence of the SM Higgs in this channel with 1.04 fb$^\mathrm{-1}$ of data.

\begin{table}[htb]
\begin{center}
\begin{tabular}{|c|crcrc|} 
\hline 
Source   & \multicolumn{2}{l}{expected events} & (stat.) &  & (sys.) \\ 
\hline
$Z$+jets &  54.4 &  $ \pm $  & 3.9 & $ \pm $ & 12.3 \\
$W$+jets &  466.7 &  $ \pm $  & 1.4 & $ \pm $ & 66.5 \\
Top-quark &  1141.8 &  $ \pm $  & 8.8 & $ \pm $ & 78.0 \\
Multijet &  193.0 &  $ \pm $  & 9.4 & $ \pm $ & 96.5 \\
$WZ$ &  16.1 &  $ \pm $  & 2.2 & $ \pm $ & 3.4 \\
$WW$ &  4.8 &  $ \pm $  & 1.1 & $ \pm $ & 1.4 \\
\hline 
Total background &  1876.8& $ \pm $ & 13.7  & $ \pm $ & 147.2 \\
Data &	1888 & & & & \\
\hline
Signal $m_\mathrm{H}=110$ GeV  &  6.72 & $ \pm $ & 0.31  & $ \pm $ & 1.20  \\
Signal $m_\mathrm{H}=115$ GeV  &  5.25 & $ \pm $ & 0.30  & $ \pm $ & 0.97  \\
Signal $m_\mathrm{H}=120$ GeV  &  4.54 & $ \pm $ & 0.25  & $ \pm $ & 0.83  \\
Signal $m_\mathrm{H}=125$ GeV  &  4.08 & $ \pm $ & 0.21  & $ \pm $ & 0.77  \\
Signal $m_\mathrm{H}=130$ GeV  &  3.28 & $ \pm $ & 0.17  & $ \pm $ & 0.62  \\
\hline 
\end{tabular}
\end{center}
\caption{Expected background events, observed events in data, and predicted signal event count in the range of 40 $<~m_\mathrm{b\bar{b}}~<$ 250 GeV. Electron and muon channels are combined. Statistical and systematic uncertainties on the expected background and signal events are shown.}
\label{tab:wh_res}
\end{table}

\begin{figure}[ht]
 \centering
 \includegraphics[width=80mm]{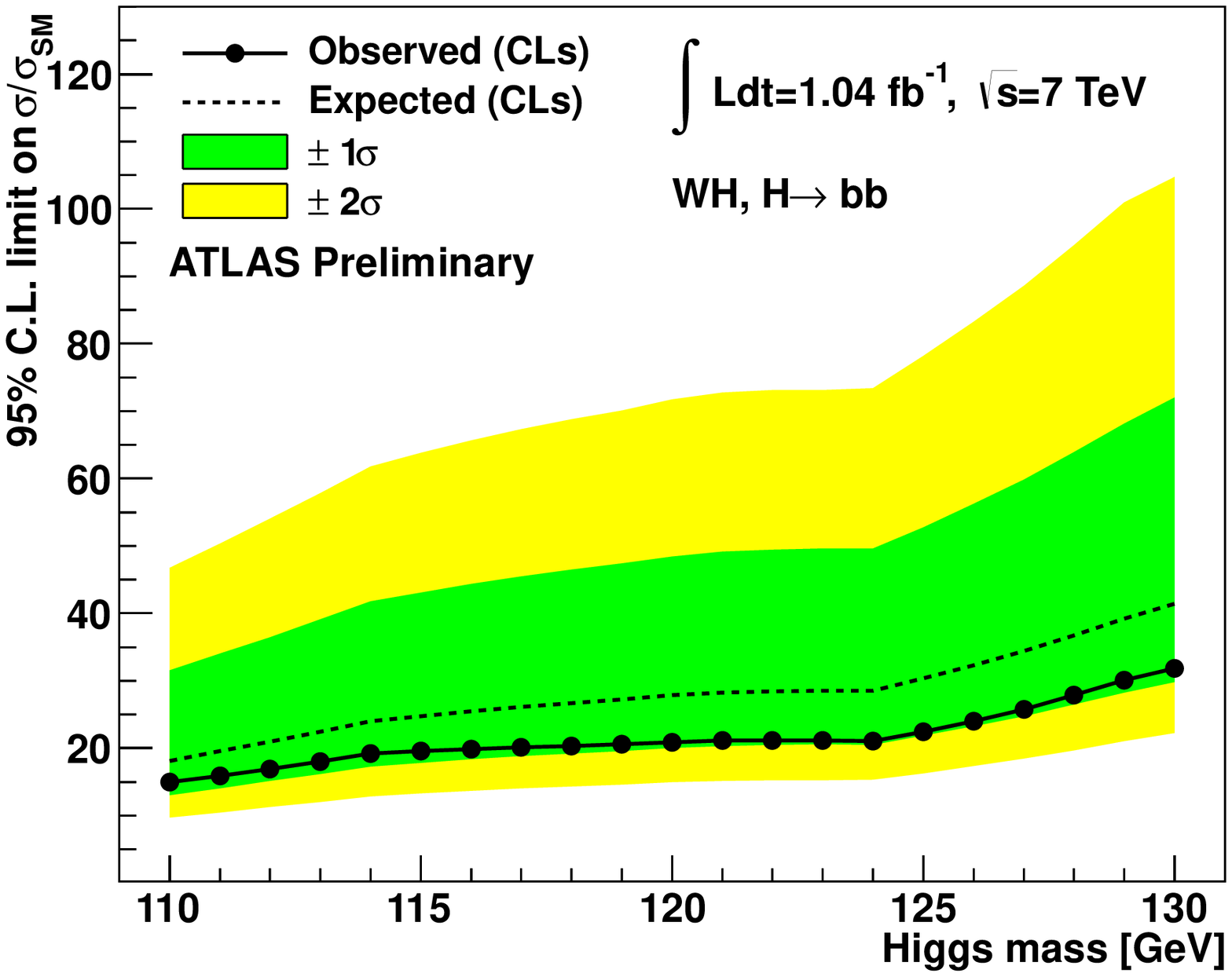}
 \includegraphics[width=80mm]{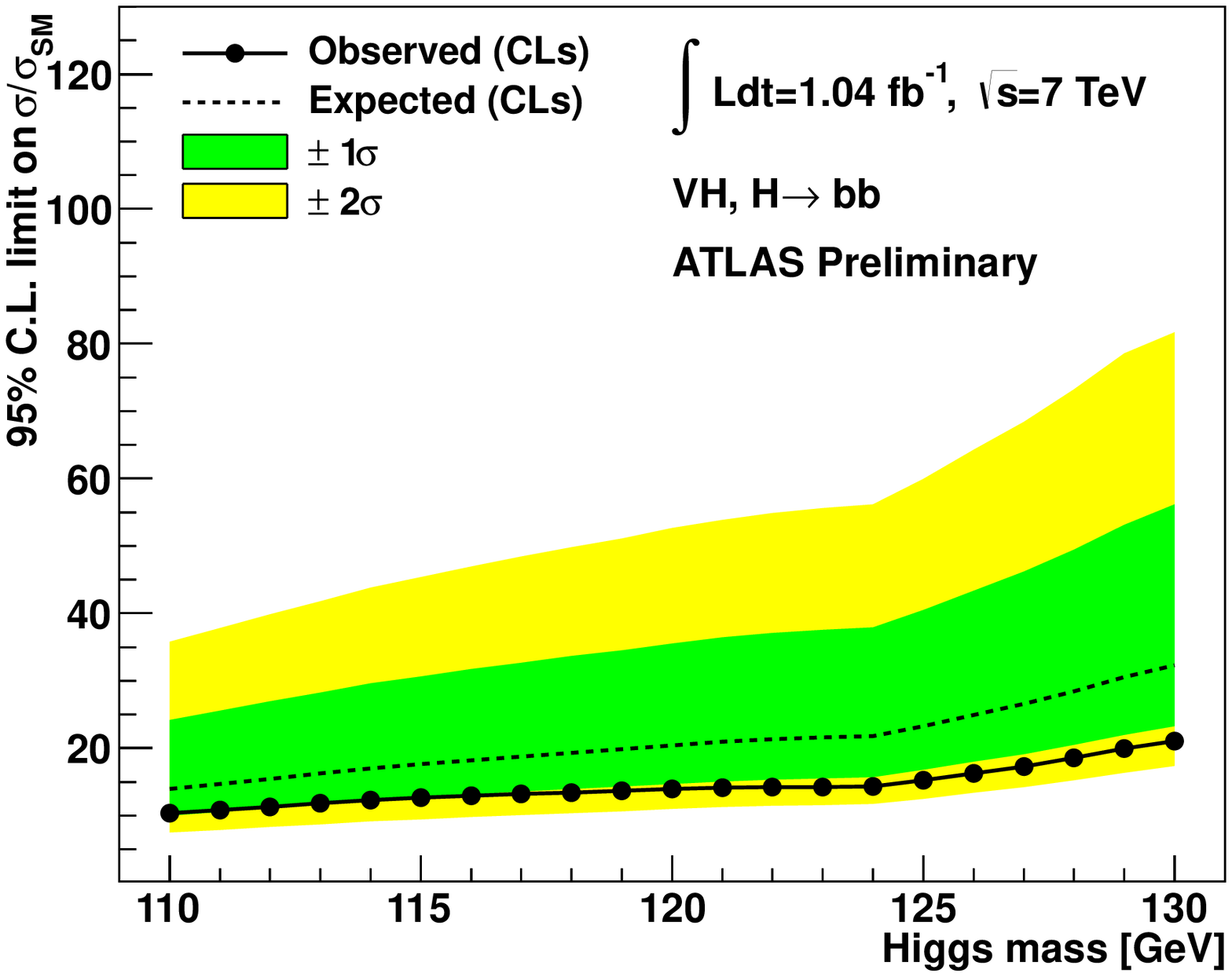}
 \caption{Expected (dashed) and observed (solid line) exclusion limits from the $WH \rightarrow l\nu bb$ channel alone \textit{(left)} and from the $WH \rightarrow l\nu bb$ and $ZH \rightarrow llbb$ channels combined \textit{(right)}.}
 \label{fig:wh_lim}
\end{figure}

Figure~\ref{fig:wh_lim} \textit{(left)} shows the exclusion limit on the standardized cross-section $\mu = \sigma/\sigma_\mathrm{SM}$ at the 95\% confidence level. The observed limits are seen to range between 15 to 30 times the SM Higgs cross-section over the Higgs mass range investigated. Figure~\ref{fig:wh_lim} \textit{(right)} shows the exclusion limit obtained by combining the $ZH \rightarrow llbb$ channel with the $WH \rightarrow l\nu bb$ channel~\cite{wh_conf}. The observed limits from the combined search range between 10 to 20 times the SM Higgs cross-section. 

\section{The $H \rightarrow WW^{(*)} \rightarrow l\nu l\nu$ channel}

$H \rightarrow WW^{(*)} \rightarrow l\nu l\nu~(l = e~\mathrm{or}~\mu)$ is a highly sensitive channel in the SM Higgs search for Higgs masses close to 2$M_\mathrm{W}$. It is the first channel at the LHC to set exclusion limits in some Higgs mass ranges, and is a very promising discovery channel. As Figure~\ref{fig:higgs_BR} shows, the Higgs branching ratio into $WW$ is close to unity at 2$M_\mathrm{W}$, and remains large for higher masses up to a TeV. For reference, the cross-section $\times$ branching ratio of the process $H \rightarrow WW^{(*)} \rightarrow l\nu l\nu$ is about 0.36 pb for $M_\mathrm{H}$ = 150 GeV.

The analysis presented here covers the range of 110 $<~m_\mathrm{H}~<$ 240 GeV. The analysis is divided into six final states based on the leptonic decay modes of the $W$ pair and the jet multiplicity: $ee$, $e\mu$, $\mu\mu$, each accompanied by either no jet or exactly 1 jet. The event selection basically involves selecting two high-$p_\mathrm{T}$ isolated leptons and large $E_\mathrm{T}^\mathrm{miss}$.

\subsection{Event selection} 
\label{sec:ww_evsel}

As in the $WH$ analysis, lowest-threshold unprescaled triggers are used. In this case, the trigger requirement depends on the final state:

\begin{itemize}

 \item $ee$: single electron trigger with 20 GeV threshold
 \item $\mu\mu$: OR of two single muon triggers, with thresholds at 18 GeV and 40 GeV\footnote{The 40 GeV threshold trigger is used to increase acceptance in the barrel region of the detector.}
 \item $e\mu$: OR of the above three triggers
 
\end{itemize}

The intial event selection, termed \textit{pre-selection}, is described below:

\begin{itemize}

 \item Exactly two leptons in the event, the higher-$p_\mathrm{T}$ ('leading') lepton having $p_\mathrm{T}~>$ 25 GeV. The lower-$p_\mathrm{T}$ ('sub-leading') lepton must have $p_\mathrm{T}~>$ 20 GeV for electrons and $p_\mathrm{T}~>$ 15 GeV for muons.
 
 \item In the $ee$ and $\mu\mu$ channels, the dilepton invariant mass $m_\mathrm{ll}~>$ 15 GeV to reject a low mass Drell-Yan background. In addition, $|m_\mathrm{ll} - m_\mathrm{Z}|~>$ 15 GeV to reject $Z$ background. Finally, the relative missing transverse energy, $E_\mathrm{T,rel}^\mathrm{miss}~>$ 40 GeV. $E_\mathrm{T,rel}^\mathrm{miss}$ is defined as:
 
  \begin{equation}
  \label{eq:metreldef}
   E_\mathrm{T,rel}^\mathrm{miss} = \left\{
   \begin{array}{l l}
    E_\mathrm{T}^\mathrm{miss}  & \textrm{ if } \Delta\phi \ge \pi/2 \\
    E_\mathrm{T}^\mathrm{miss} \cdot \sin\Delta\phi  & \textrm{ if } \Delta\phi < \pi/2 \\ \end{array},\right.
  \end{equation}

  where $\Delta\phi$ is the absolute value of the difference in the azimuthal angle $\phi$ between the $E_\mathrm{T}^\mathrm{miss}$ and the nearest lepton with $p_\mathrm{T}~>$ 15 GeV or jet with $p_\mathrm{T}~>$ 25 GeV. 
  
 \item In the $e\mu$ channel, $m_\mathrm{ll}~>$ 10 GeV and $E_\mathrm{T,rel}^\mathrm{miss}~>$ 25 GeV.
 
\end{itemize}

\begin{figure}[ht]
 \centering
 \includegraphics[width=80mm]{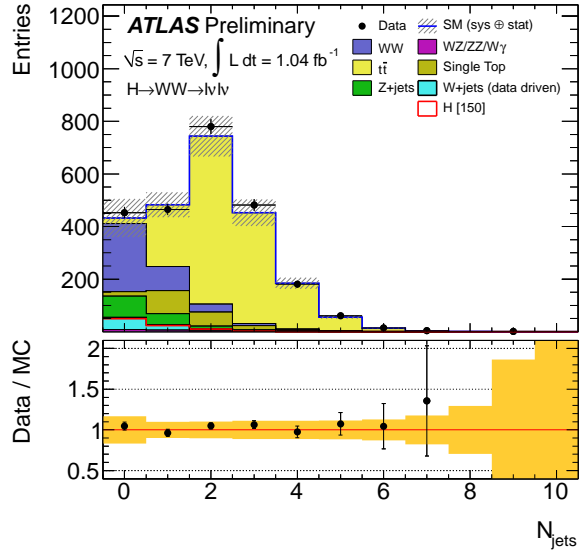}
 \caption{Jet multiplicity after the cut on $E_\mathrm{T,rel}^\mathrm{miss}$. The bin-by-bin ratio between data and the background expectation is also shown, with the overall systematic uncertainty in the normalizations indicated by the yellow band.}
 \label{fig:ww_njet}
\end{figure}

Following the pre-selection, the events are divided into 0-jet and 1-jet bins. For this purpose, each jet is required to have $p_\mathrm{T}~>$ 25 GeV and $|\eta|~<$ 4.5. Figure~\ref{fig:ww_njet} shows the jet multiplicity distribution after the $E_\mathrm{T,rel}^\mathrm{miss}$ cut for data, for MC signal and for various background predictions.

At this stage, different cuts are applied to events depending on the jet bin. In the 0-jet bin, the transverse momentum of the dilepton system, $p_\mathrm{T}^\mathrm{ll}$, must be greater than 30 GeV. Then a series of topological cuts is applied depending on the Higgs mass:

\begin{itemize}

 \item $m_\mathrm{H}~<$ 170 GeV: $m_\mathrm{ll}~<$ 50 GeV; the dilepton opening angle, $\Delta\phi_\mathrm{ll}~<$ 1.3.
 
 \item $m_\mathrm{H}~>=$ 170 GeV: $m_\mathrm{ll}~<$ 65 GeV; $\Delta\phi~<$ 1.8.
 
 \item The transverse mass $m_\mathrm{T}$ must satisfy  $0.75\times m_\mathrm{H} < m_\mathrm{T} < m_\mathrm{H}$, where
 
  \begin{equation}
  \label{eq:mT}
   m_{\rm T}=\sqrt{(E_{\rm T}^\mathrm{\ell\ell}+E_{T}^\mathrm{miss})^\mathrm{2}-({\bf P}_{\rm T}^\mathrm{\ell\ell}+{\bf P}_{\rm T}^{\rm miss})^\mathrm{2}},
  \end{equation}
 
  with $E_{\rm T}^\mathrm{\ell\ell}=\sqrt{({\bf P}_{\rm T}^\mathrm{\ell\ell})^\mathrm{2}+m_{\ell\ell}^\mathrm{2}}$, $|{\bf P}_{\rm T}^{\rm miss}|=E_{T}^\mathrm{miss}$ and ${\bf P}_{\rm T}^\mathrm{\ell\ell}$ is the transverse momentum of the dilepton system.
 
 \end{itemize}
 
In the 1-jet channel, a $b$-jet veto is applied to reduce top background. The magnitude of the total $p_\mathrm{T}$ of the Higgs boson and jet system, defined as ${\bf P}_{\rm T}^{\rm tot}={\bf P}_{\rm T}^{\rm l1}+{\bf P}_{\rm T}^{\rm l2}+{\bf P}_{\rm T}^{\rm j}+{\bf P}_{\rm T}^{\rm miss}$, is required to be less than 30 GeV. Finally, the same topological selection as in the 0-jet case is applied.

\subsection{Background processes}

\textit{WW}: Continuum $WW$ production is the largest background in the 0-jet channel. A dedicated control region in data is used to derive the $WW$  normalization in the signal region. This control region is defined using the same pre-selection as for the signal region, but requiring $m_\mathrm{ll}~>$ 80 GeV and removing the topological cuts. Table~\ref{tab:ww_wwctrl} shows the expected number of signal and background events in the $WW$ control region, together with the observed number of events in data.  

\begin{table}[tbp]
 \begin{center}
 \vspace*{0.2cm}
 {\scriptsize
 \begin{tabular}{| c | c| c| c| c| c| c| c| c | c |}
 \hline
  & Signal & $WW$ & $W$+jets & $Z/\gamma^*$+jets & $t\bar{t}$ & $tW/tb/tqb$ & $WZ/ZZ/W\gamma$ & Total Bkg. & Observed \\
 \hline
 $ee+e\mu+\mu\mu$ & $0.73 \pm 0.16$ & $111 \pm 15$ & $13 \pm 7$ & $2 \pm 3$ & $10 \pm 4$ & $8 \pm 2$ & $4.3 \pm 1.2$ & $150 \pm 30$ & 153 \\ 
 \hline
 $ee$ & $0.01 \pm 0.07$ & $13 \pm 2$ & $2 \pm 2$ & $0.2 \pm 1.7$ & $1.4 \pm 1.6$ & $0.5 \pm 1.0$ & $0 \pm 2$ & $17 \pm 5$ & 28 \\ 
 $e\mu$ & $0.73 \pm 0.17$ & $76 \pm 10$ & $9 \pm 6$ & $< 0.01$ & $6 \pm 3$ & $4.6 \pm 1.4$ & $3 \pm 2$ & $100 \pm 20$ & 98 \\ 
 $\mu\mu$ & $< 0.14$ & $23 \pm 3$ & $1.8 \pm 1.2$ & $1.9 \pm 1.2$ & $2.2 \pm 1.7$ & $2.6 \pm 1.1$ & $1 \pm 2$ & $32 \pm 6$ & 27 \\ 
 \hline
 \end{tabular}
 }
 \end{center}
 \caption{Expected and observed event numbers after all cuts in the $WW$ control region for the 0-jet channel in 1.04 fb$^{-1}$ of data. The signal numbers correspond to $m_\mathrm{H} = 150$~GeV.}
 \label{tab:ww_wwctrl}
\end{table}

\textit{Top}: $t\bar{t}$ and single top constitute the largest background in the 1-jet channel. The signal region normalization for these backgrounds is determined from a control region in data defined by reversing the $b$-jet veto and removing the topological cuts. 

\textit{W+jets}: This background is estimated using a fully data-driven method. A control sample in data is obtained by requiring one lepton in the events to pass the full set of selection criteria listed in Section~\ref{sec:ww_evsel}, while the other lepton must pass a looser set of criteria. A normalization \textit{fake factor} is extracted from a dijet data sample, which is then applied to the control sample to estimate the $W+jets$ contamination in the signal region. 

\textit{Z+jets}: The \textit{Z+jets} background prediction is taken from MC, normalized by a $E_\mathrm{T}^\mathrm{miss}$ mismodelling factor obtained from a data/MC comparison in two control regions.

\textit{WZ, ZZ, W$\gamma$}: These backgrounds are small, and are estimated from MC.

\subsection{Results: 0-jet channel}

\begin{figure}[ht]
 \centering
 \includegraphics[width=75mm]{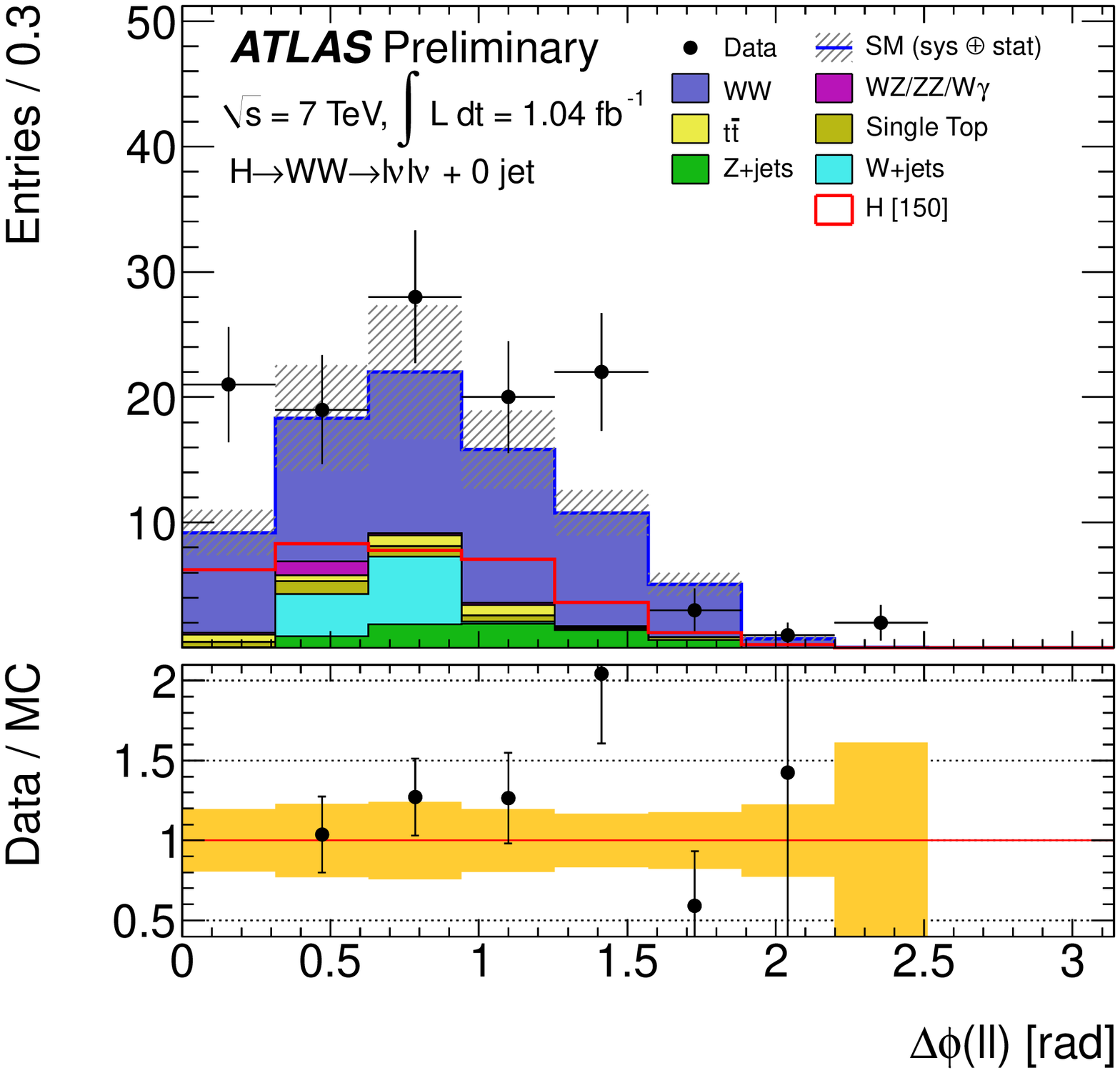}
 \includegraphics[width=75mm]{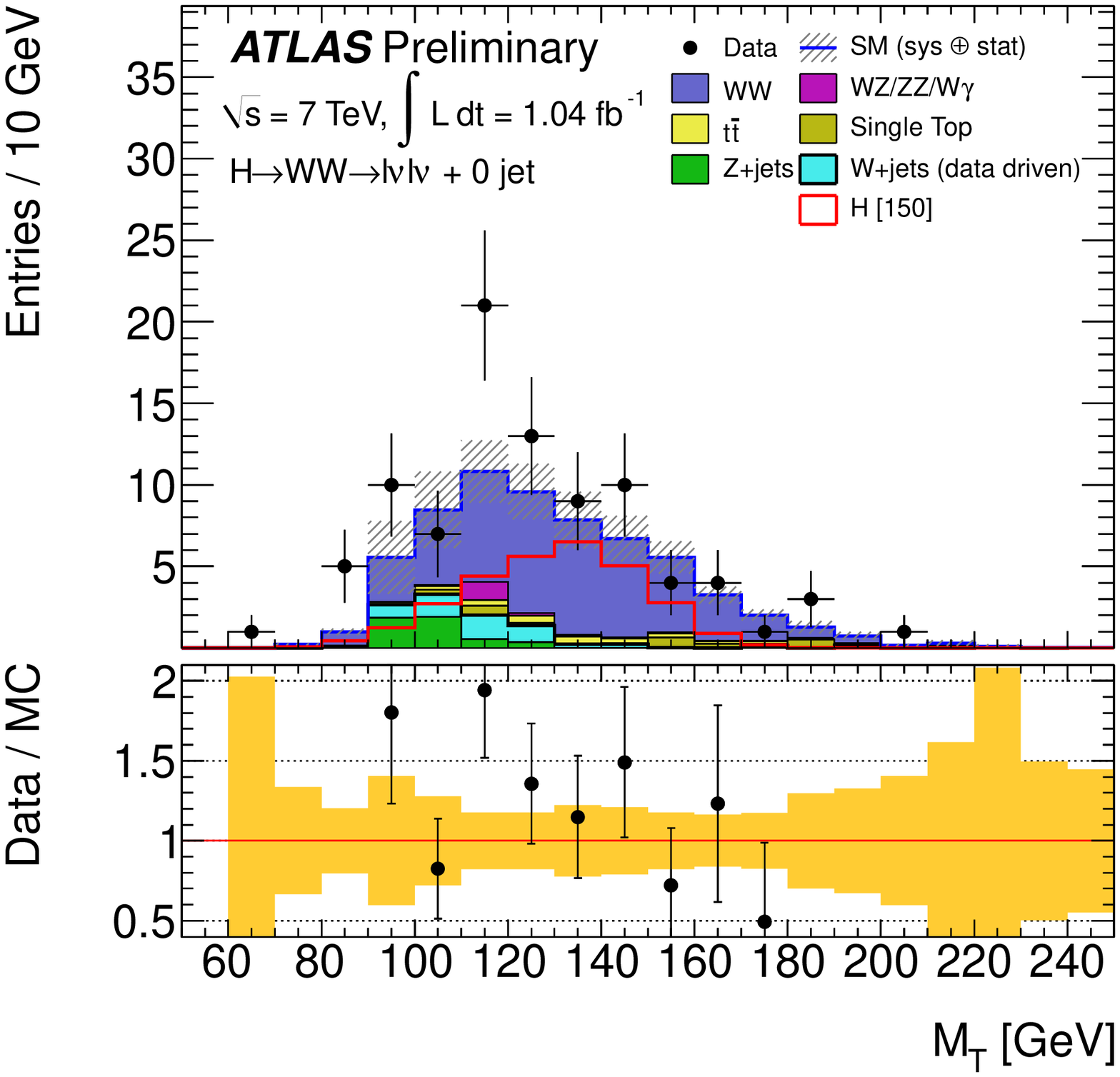}
 \caption{Distributions of the dilepton opening angle $\Delta\phi_\mathrm{ll}$ \textit{(left)} and of the transverse mass $m_\mathrm{T}$ \textit{(right)} in the 0-jet channel after the cut on $m_\mathrm{ll}$. The bin-by-bin ratio between data and the background expectation is also shown, with the overall systematic uncertainty indicated by the yellow band.}
 \label{fig:ww_0jetkin}
\end{figure}

Table~\ref{tab:ww_0jet} shows the expected number of signal and background events, as well as the observed events in data, after applying each 0-jet selection. We observe a small excess in data in the $\mu\mu$ channel after all cuts. Figure~\ref{fig:ww_0jetkin} shows the $\Delta\phi_\mathrm{ll}$ and the $m_\mathrm{T}$ distributions in data and for signal and background expectations. The data seems to be systematically higher than the expectation. 

\begin{table}[htbp]
\begin{center}
\vspace*{0.2cm}
{\scriptsize
\begin{tabular}{| c | c |c |c |c |c |c |c |c | c |}
\hline
Selection & Signal & $WW$ & $W$+jets & $Z/\gamma^*$+jets & $t\bar{t}$ & $tW/tb/tqb$ & $WZ/ZZ/W\gamma$ & Total Bkg. & Observed\\ 
\hline
Jet Veto & $50 \pm 11$ & $260 \pm 30$ & $46 \pm 17$ & $80 \pm 70$ & $22 \pm 8$ & $17 \pm 4$ & $7.8 \pm 1.5$ & $430 \pm 100$ & 453 \\ 
$|{\bf P}_{\rm T}^{\ell\ell}|>30$ GeV & $48 \pm 10$ & $230 \pm 20$ & $38 \pm 14$ & $15 \pm 6$ & $19 \pm 7$ & $16 \pm 4$ & $7.3 \pm 1.4$ & $330 \pm 50$ & 371 \\ 
$m_{\ell\ell} < 50$ GeV & $34 \pm 7$ & $59 \pm 8$ & $11 \pm 3$ & $7 \pm 4$ & $2.7 \pm 1.8$ & $2.8 \pm 0.8$ & $0.9 \pm 0.3$ & $83 \pm 11$ & 116 \\ 
$\Delta\Phi_{\ell\ell} < 1.3$ & $30 \pm 7$ & $46 \pm 6$ & $5.8 \pm 1.8$ & $5 \pm 3$ & $2.7 \pm 1.7$ & $2.8 \pm 0.8$ & $0.8 \pm 0.2$ & $63 \pm 9$ & 89 \\ 
$0.75\times m_{H}< m_{\rm T}< m_{H}$ & $21 \pm 4$ & $26 \pm 3$ & $2.9 \pm 0.9$ & $1 \pm 2$ & $1.6 \pm 1.2$ & $0.7 \pm 0.4$ & $0.6 \pm 0.2$ & $33 \pm 5$ & 49 \\ 
\hline
$ee$ & $3.1 \pm 0.7$ & $3.7 \pm 0.7$ & $0.5 \pm 0.2$ & $0.4 \pm 0.6$ & $0.0 \pm 0.6$ & $0.0 \pm 0.2$ & $0.05 \pm 0.19$ &  $4.7 \pm 1.2$ & 7 \\ 
$e\mu$ & $11 \pm 2$ & $13.4 \pm 1.9$ & $1.7 \pm 0.7$ & $< 0.01$ & $1.1 \pm 0.8$ & $0.4 \pm 0.3$ & $0.4 \pm 0.3$ & $17 \pm 2$ & 21 \\ 
$\mu\mu$ & $6.9 \pm 1.5$ & $8.8 \pm 1.3$ & $0.7 \pm 0.5$ & $0.5 \pm 2.0$ & $0.4 \pm 0.8$ & $0.3 \pm 0.3$ & $0.18 \pm 0.19$ & $11 \pm 3$ & 21 \\ 
\hline
\end{tabular}
}
\end{center}
\caption{Expected and observed event numbers after applying each cut for the 0-jet channel in 1.04 fb$^\mathrm{-1}$ of data. The signal numbers correspond to $m_\mathrm{H} = 150$~GeV. The $W$+jets background expectation is purely data-driven. The uncertainties shown include both statistical and systematic components.}
\label{tab:ww_0jet}
\end{table}

\subsection{Results: 1-jet channel}

\begin{figure}[ht]
 \centering
 \includegraphics[width=77mm]{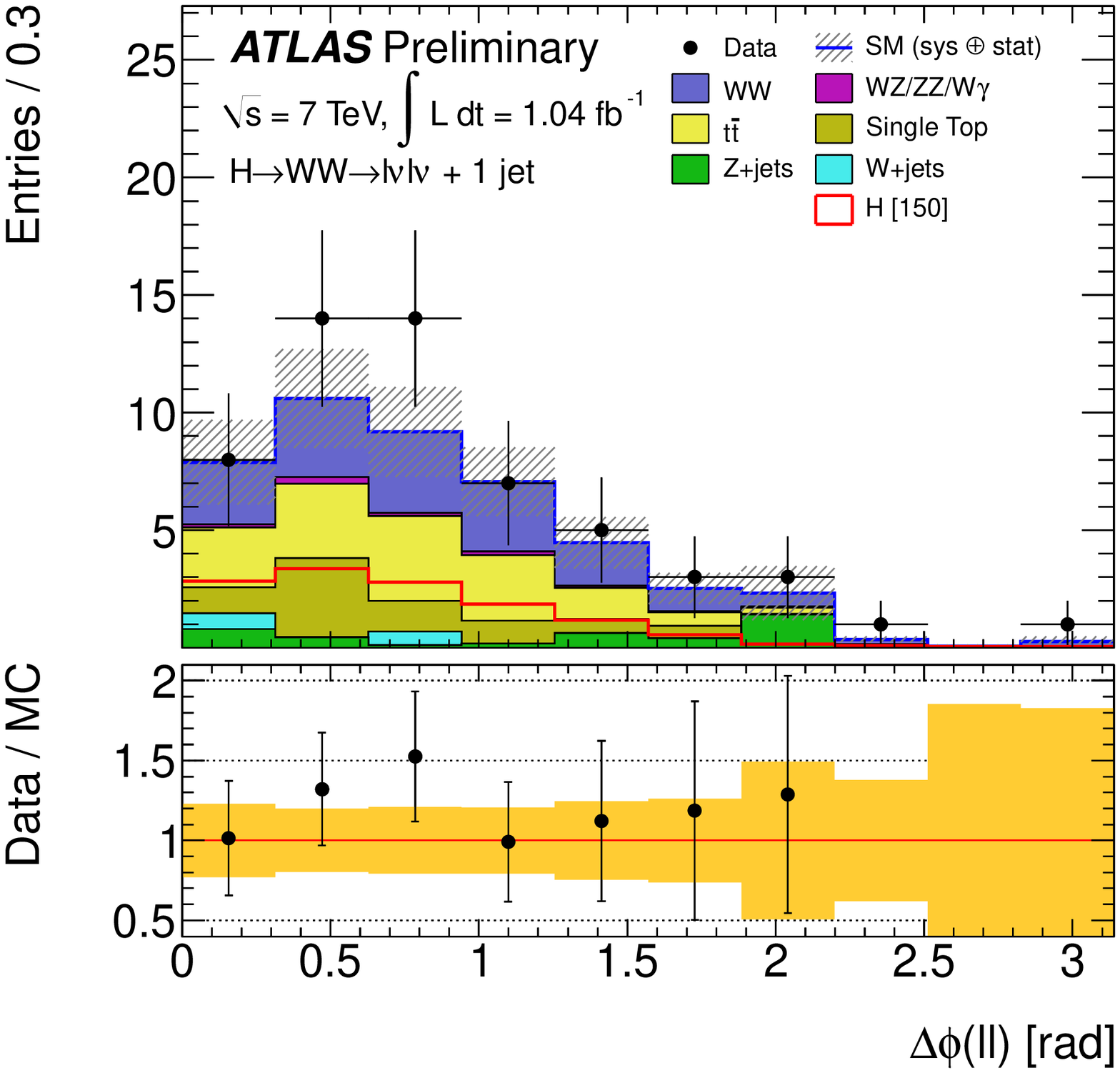}
 \includegraphics[width=77mm]{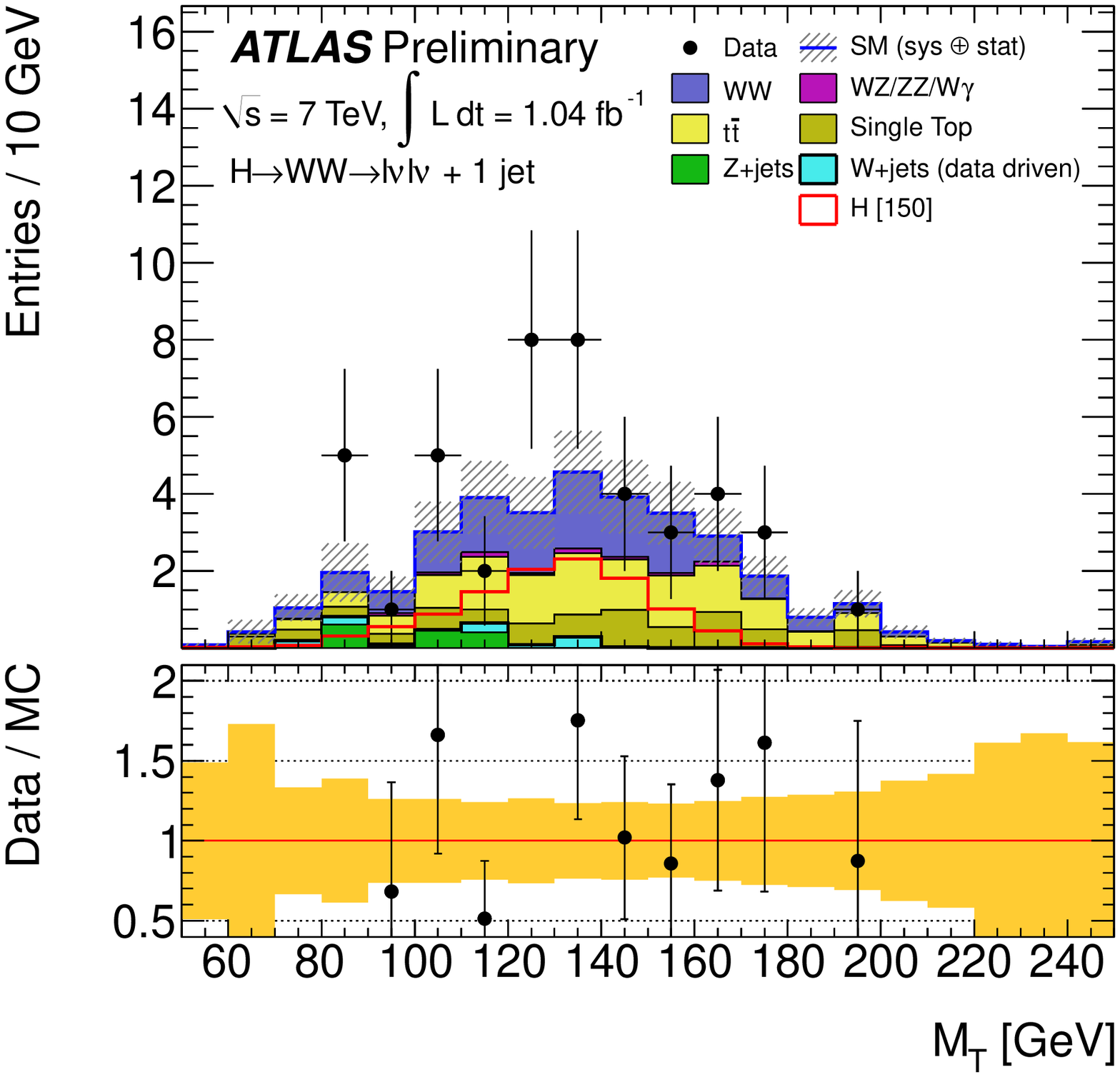}
 \caption{\textit{Left}: Distributions of the dilepton opening angle $\Delta\phi_\mathrm{ll}$  and of the transverse mass $m_\mathrm{T}$ \textit{(right)} in the 1-jet channel after the cut on $m_\mathrm{ll}$. The bin-by-bin ratio between data and the background expectation is also shown, with the overall systematic uncertainty indicated by the yellow band.}
 \label{fig:ww_1jetkin}
\end{figure}

Table~\ref{tab:ww_1jet} shows the expected number of signal and background events and the observed events in data after applying each 1-jet selection. As in the 0-jet case, we see a small excess in data in the $\mu\mu$ channel after all cuts. Figure~\ref{fig:ww_1jetkin} shows the $\Delta\phi_\mathrm{ll}$ and the $m_\mathrm{T}$ distributions in data and for signal and background expectations. The same trend is seen in the data-MC comparison as in the 0-jet case. 

\begin{table}[htbp]
\begin{center}
{\scriptsize
\begin{tabular}{|c| c|c|c|c|c|c|c|c | c|}\hline
Selection & Signal & $WW$ & $W$+jets & $Z/\gamma^*$+jets & $t\bar{t}$ &  $tW/tb/tqb$ & $WZ/ZZ/W\gamma$ & Total Bkg. & Observed \\ 
\hline
1 jet & $23 \pm 4$ & $92 \pm 9$ & $20 \pm 10$ & $40 \pm 30$ & $240 \pm 60$ & $88 \pm 13$ & $6.2 \pm 0.8$ & $490 \pm 70$ & 465 \\ 
$b$-jet veto & $23 \pm 4$ & $91 \pm 9$ & $19 \pm 10$ & $40 \pm 30$ & $140 \pm 40$ & $45 \pm 7$ & $6.1 \pm 0.8$ & $340 \pm 50$ & 333 \\ 
$|{\bf P}_{T}^{\rm tot}|<30$ GeV & $19 \pm 3$ & $76 \pm 8$ & $9 \pm 5$ & $25 \pm 19$ & $80 \pm 20$ & $35 \pm 6$ & $4.1 \pm 0.5$ & $230 \pm 40$ & 221 \\ 
$Z\to\tau\tau$ veto & $19 \pm 4$ & $74 \pm 8$ & $9 \pm 5$ & $20 \pm 10$ & $80 \pm 19$ & $33 \pm 5$ & $4.0 \pm 0.7$ & $220 \pm 17$ & 212 \\ 
$m_{\ell\ell} < 50$ GeV & $13 \pm 3$ & $16 \pm 3$ & $1.2 \pm 0.5$ & $3.4 \pm 1.6$ & $12 \pm 4$ & $7.2 \pm 1.7$ & $0.9 \pm 0.2$ & $41 \pm 5$ & 56 \\ 
$\Delta\Phi_{\ell\ell} < 1.3$ & $11 \pm 2$ & $13 \pm 2$ & $1.0 \pm 0.5$ & $1.5 \pm 1.2$ & $11 \pm 4$ & $6.3 \pm 1.5$ & $0.74 \pm 0.20$ & $33 \pm 5$ & 44 \\ 
$0.75\times m_{H}< m_{\rm T}< m_{H}$ & $7.2 \pm 1.6$ & $6.2 \pm 1.3$ & $0.5 \pm 0.9$ & $0.4 \pm 0.6$ & $4.9 \pm 1.7$ & $2.3 \pm 0.7$ & $0.34 \pm 0.16$ & $15 \pm 3$ & 21 \\ 
\hline
$ee$ & $0.9 \pm 0.3$ & $0.8 \pm 0.3$ & $0.08 \pm 0.04$ & $0.0 \pm 0.4$ & $0.8 \pm 1.0$ & $0.2 \pm 0.4$ & $0.06 \pm 0.08$ & $2.0 \pm 1.2$ & 4 \\ 
$e\mu$ & $4.0 \pm 0.9$ & $3.5 \pm 0.8$ & $0.4 \pm 0.2$ & $0.4 \pm 0.7$ & $3.1 \pm 1.3$ & $1.2 \pm 0.6$ & $0.24 \pm 0.13$ & $8.8 \pm 1.9$ & 8 \\ 
$\mu\mu$ & $2.3 \pm 0.5$ & $1.9 \pm 0.4$ & $0.0 \pm 0.8$ & $0.0 \pm 0.4$ & $1.1 \pm 1.1$ & $0.8 \pm 0.7$ & $0.04 \pm 0.07$ & $3.9 \pm 1.7$ & 9 \\ 
\hline
\end{tabular}
}
\end{center}
\caption{Expected and observed event numbers after applying each cut for the 1-jet channel in 1.04 fb$^\mathrm{-1}$ of data. The signal numbers correspond to $m_\mathrm{H} = 150$~GeV. The $W$+jets background expectation is purely data-driven. The uncertainties shown include both statistical and systematic components.}
\label{tab:ww_1jet}
\end{table}

\subsection{Exclusion limits}

\begin{figure}[ht]
 \centering
 \includegraphics[width=80mm]{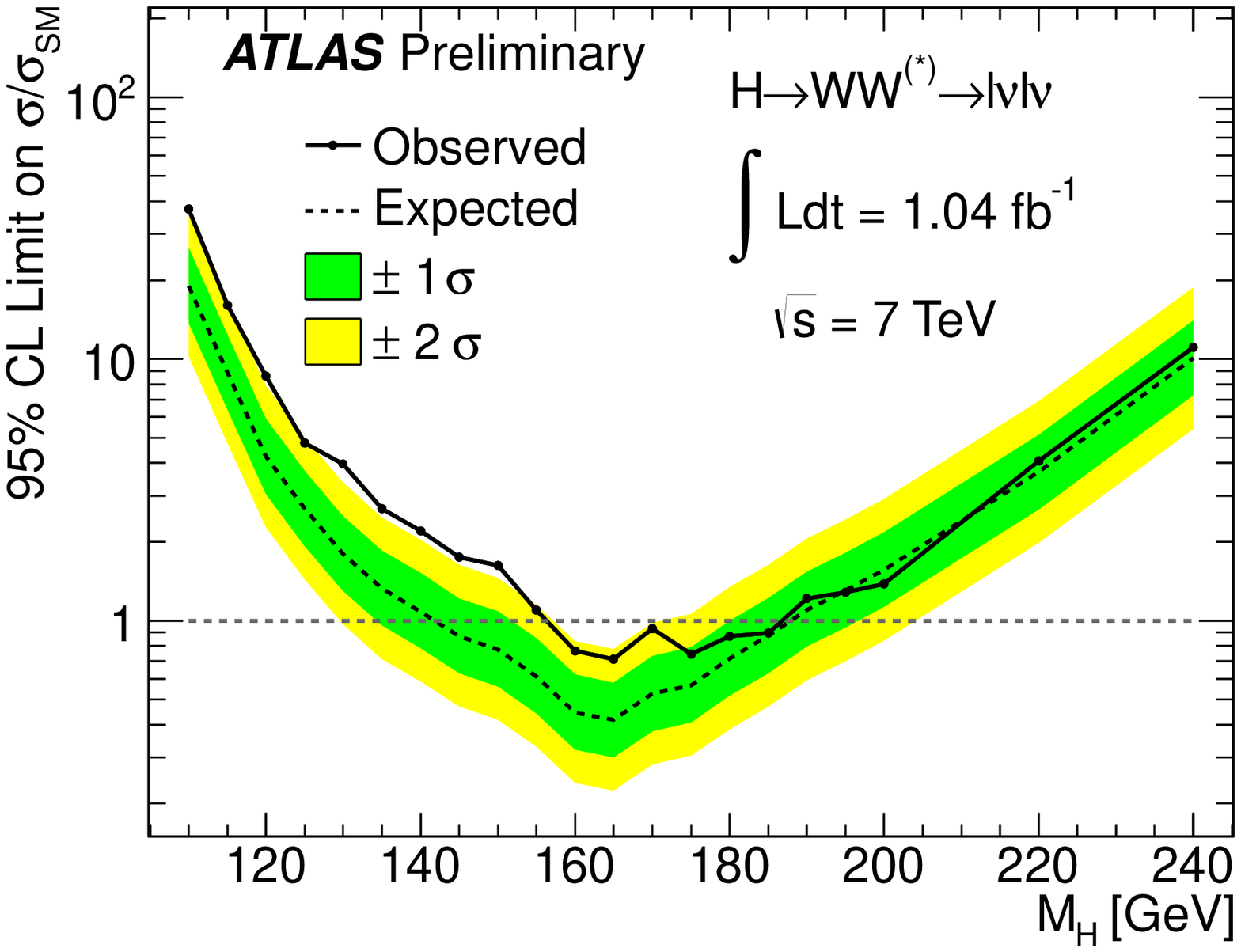}
 \includegraphics[width=80mm]{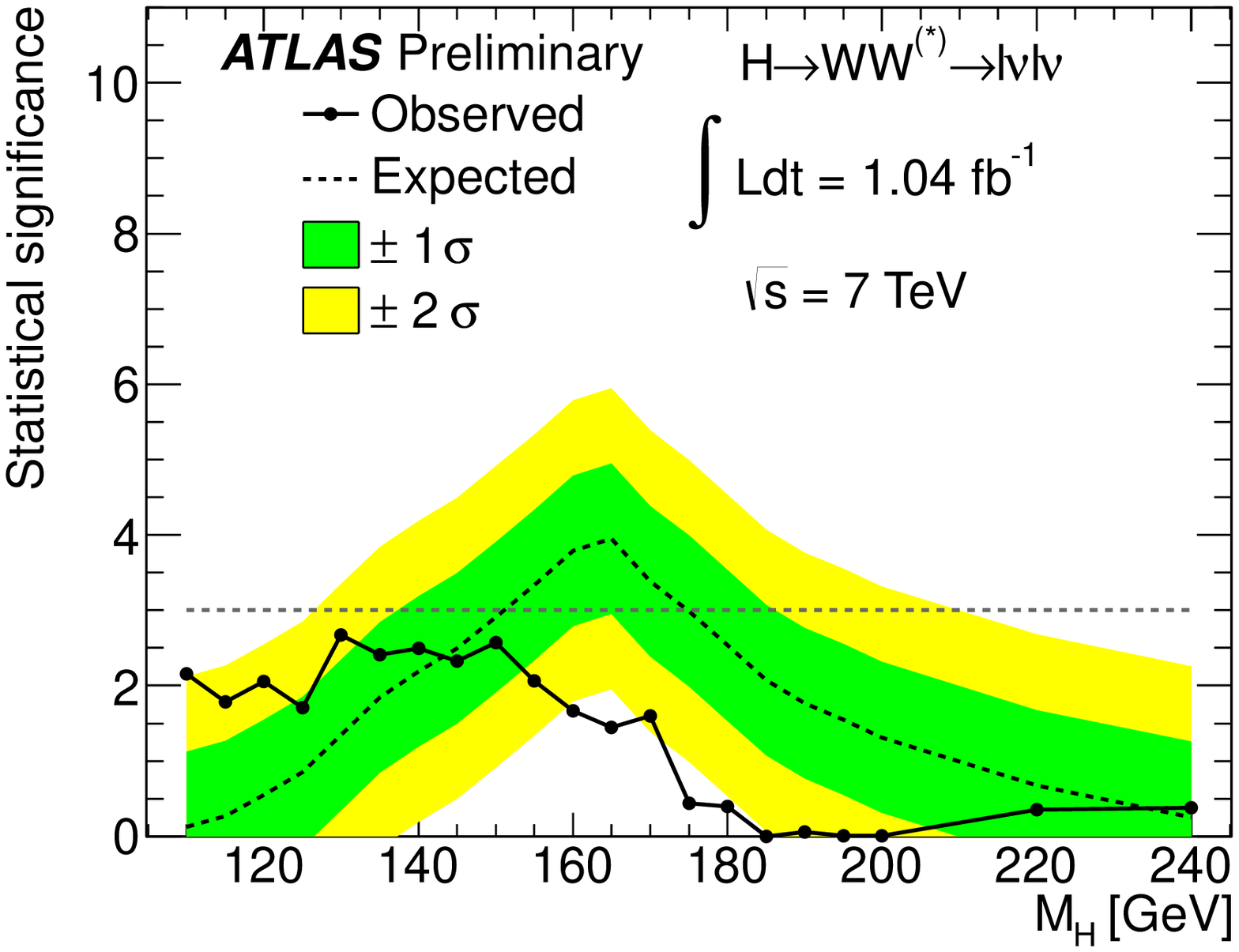}
 \caption{\textit{Left}: Expected (dashed) and observed (solid line) exclusion limits from the $H \rightarrow WW^{(*)} \rightarrow l\nu l\nu$ channel; \textit{right}: the expected (dashed) and observed (solid) signal significances for a SM Higgs boson production vs $m_\mathrm{H}$.}
 \label{fig:ww_lim}
\end{figure}

Figure~\ref{fig:ww_lim} \textit{(left)} shows the exclusion limit on the standardized cross-section $\mu = \sigma/\sigma_\mathrm{SM}$ from the $H \rightarrow WW^{(*)} \rightarrow l\nu l\nu$ channel. The Standard Model Higgs is excluded in the mass range of 158 $<~m_\mathrm{H}~<$ 186 GeV at the 95\% confidence level, the expected limit being 142 $<~m_\mathrm{H}~<$ 186 GeV. Figure~\ref{fig:ww_lim} \textit{(right)} shows the signal significances as functions of the Higgs mass. In the range 126 $<~m_\mathrm{H}~<$ 158 GeV, an excess in data corresponding to larger than 2$\sigma$ significance is observed. The largest excess is observed at $m_\mathrm{H}\approx$ 130 GeV, and corresponds to about 2.7$\sigma$.

\section{Conclusion}

We have presented results of the SM Higgs search by ATLAS in the $WH \rightarrow l\nu bb$ and $H \rightarrow WW^{(*)} \rightarrow l\nu l\nu$ channels using 1.04 fb$^\mathrm{-1}$ of data collected in 2011. The $WH \rightarrow l\nu bb$ channel sets limits on the Higgs production cross-section that are 15-30 times the Standard Model cross-section in the range 110 $<~m_\mathrm{H}~<$ 130 GeV, while the $H \rightarrow WW^{(*)} \rightarrow l\nu l\nu$ channel excludes the SM Higgs in the range of 158 $<~m_\mathrm{H}~<$ 186 GeV at the 95\% confidence level.

Expected improvements to both analyses are going to include multivariate approaches, reduced systematics uncertainties and a boosted Higgs analysis in the $H \rightarrow b\bar{b}$ channel.

\bigskip 

\end{document}